\documentclass{article}

\usepackage{PRIMEarxiv}

\usepackage[utf8]{inputenc} 
\usepackage[T1]{fontenc}    
\usepackage{hyperref}       
\usepackage{url}            
\usepackage{booktabs}       
\usepackage{amsfonts}       
\usepackage{amsmath}
\usepackage{nicefrac}       
\usepackage{microtype}      
\usepackage[round]{natbib}
\usepackage{lipsum}
\usepackage{fancyhdr}       
\usepackage{graphicx}       
\graphicspath{{media/}}     
\usepackage{xcolor}         

\title{Recurrent neural networks with more flexible memory: better predictions than rough volatility}

\author{
  Damien Challet$^1$ and Vincent Ragel$^{1,2}$ \\
  $^1$ Université Paris Saclay, CentraleSupélec, Laboratoire MICS, \\
  91190 Gif-sur-Yvette, France\\
  $^2$ BNP Paribas, 20 boulevard des Italiens, 75008 Paris, France\\
  \texttt{{damien.challet,vincent.ragel}@centralesupelec.fr} 
}

\begin{document}
\maketitle


\begin{abstract}
We extend recurrent neural networks to include several flexible timescales for each dimension of their output, which mechanically improves their abilities to account for processes with long memory or with highly disparate time scales. We compare the ability of vanilla and extended long short term memory networks (LSTMs) to predict asset price volatility, known to have a long memory. Generally, the number of epochs needed to train extended LSTMs is divided by two, while the variation of validation and test losses among models with the same hyperparameters is much smaller. We also show that the model with the smallest validation loss systemically outperforms rough volatility predictions by about 20\% when trained and tested on a dataset with multiple time series.
\end{abstract}

\keywords{Time series \and Long memory \and Recurrent Neural Networks \and Rough Volatility \and Volatility modelling}

\section{Introduction}

Some time series in Nature have a very long memory \citep{robinson2003time}: fluid turbulence \citep{resagk2006oscillations}, asset price volatility \citep{cont2001empirical} and tick-by-tick events in financial markets \citep{challet2001analyzing,lillo2004long}. From a modelling point of view, this means that the current value of an observable of interest depends on the past by a convolution of itself with a long-tailed kernel. 

Deep learning tackles past dependence in time series with recurrent neural networks (RNNs). These networks are in essence moving averages of nonlinear functions of the inputs and learn the parameters of these averages and functions. Provided that they are sufficiently large, these networks can approximate long-tailed kernels in a satisfactory way, and are of course able to account for more complex problems than a simple linear convolution. Yet, their flexibility may prevent them to learn quickly and efficiently the long memory of  time series. Several solutions exist: either one pre-filters the data by computing statistics at with various time scales and use them as inputs to RNNs in the same spirit as multi-time scale volatility modelling \citep{zumbach2001heterogeneous,corsi2009simple}, see e.g. \citet{KIM201825}, or one extends the neural networks so as to improve their abilities. For example,  \citet{zhao2020rnn} adds delay operators, taking inspiration from the ARIMA processes, to the states of recurrent neural networks, while \citet{ohno2021recurrent} modifies the update equation of the network output so that its dynamics mimics that of a variable with a long memory. In both cases, the time dependence structure is enforced by hand in the dynamics of such networks. 

Here, we propose a flexible and parsimonious way to extend the long-memory abilities of recurrent neural networks by using an old trick for approximating long-memory kernels with exponential functions, which helps recurrent neural networks learn faster and better time series with long memory.

Our main contributions are: (i) we introduce RNNs with several multiple flexible time scales for each dimension of the output; (ii) we show that learning to predict time series with long-memory (asset price volatility) is faster and more reliably good with more flexible time scales (iii) rough volatility predictions can be beaten by training a fair number of recurrent neural networks and only using the one with the best validation loss.

\section{Methods}

Let time series $y_t$ be of interest. Its moving average can be written
\begin{equation}
    \tilde{y}_t=\int_{-\infty}^t K(t-t')y_t'dt',
\end{equation}
where $K$ is a kernel. In a discrete time context,
\begin{equation}
        \tilde{y}_t=\sum_{-\infty}^t K(t-t')y_t'.
\end{equation}
When the process is Markovian, its kernel $K(x)\simeq e^{-x/\tau_0}$ for large $x$, where $\tau_0$ is the slowest timescale at which the process forgets its past. In this case, one can write $y_t$ in a recursive way
\begin{equation}\label{eq:EMA}
    \tilde{y}_t=\tilde{y}_{t-1}(1-\lambda)+\lambda y_t,
\end{equation}
where $\lambda\simeq 1/\tau_0$; $\tilde{y}_t$ is then an exponentially moving average (EMA) of $y_t$.

Long memory processes however, have a kernel that decreases at least as slowly as a power-law. In turn, power-laws can be approximated by a sum of exponential functions: naively, if $K(x)=x^{-\alpha}$, one writes
\begin{equation}\label{eq:K_exp}
    K(x)\propto \sum_{i=1}^\infty w_i \exp(-x/\tau_i)
\end{equation}
with $w_i\propto (1/c^\alpha)^i$ and $\tau_i=c^i$ for a well-chosen constant $c$: one covers the $x$ space in a geometric way and the weights $w_i$ account for the power-law decreasing nature of $K(x)$. This rough approach works well and is widespread. \cite{bochud2007optimal} propose a more refined method to determine how many exponential functions one needs to approximate optimally $K$ and how to compute $w_i$ for a given $\alpha$ and for a given range of $x$ over which the kernel has to be  approximated by a sum of exponential functions (e.g. $x\in[1,1000]$). For example, one needs about 4 exponential functions to approximate 3 decades). 

Writing down the update equations of well-known recurrent neural network architectures makes it clear that they use exponentially moving averages with a single time scale for each output dimension. For example, Gate Recurrent Units (GRU) \cite{cho2014learning} transform the input vector $x_t$ into a vector of timescales $\lambda_t$ defined as

\begin{equation}
\lambda_t=\sigma(W_\lambda x_t +U_\lambda c_{t-1} + b_\lambda )
\end{equation}

which is then used in the update of the output $c_t$
\begin{equation}\label{eq:GRU_EMA}
c_t=c_{t-1}\odot (1-\lambda)+\lambda \odot \tilde c_t,
\end{equation}
where the update $\tilde c_t$ is also computed from the input with learned weights, i.e.
\begin{align}\label{eq:c_t_update}
\tilde c_t &= \sigma_c(W_c x_t + U_c ( c_{t-1} \odot r_{t})  + b_c) \\
r_{t} &= \sigma_r(W_r x_t + U_r c_{t-1} + b_r)
\end{align}
for a non-linear functions  $\sigma_c$ and $\sigma_r$, $\odot$ is the element-wise (Hadamar) product and $r_{t}$ the reset gate which modifies the value of $c_{t-1}$ when computing $\tilde c_t$. By design, GRUs can only compute exponentially moving averages of  $\tilde c_t$, although they possess the interesting ability to learn both $\lambda_t$ and the update $\tilde c_t$ as a function of their inputs.  It is straightforward to extend GRUs to an arbitrary number of timescales $n$ by using $n$ $\tilde c^{(k)}_t $, $k=1,\cdots,n$ and 
\begin{align}\label{eq:lambda_c_k}
\lambda^{(k)}_t&=\sigma(W^{(k)}_\lambda x_t +U^{(k)}_\lambda c^{(k)}_{t-1} + b_\lambda^{(k)} )\\
c_t^{(k)}&=c_{t-1}^{(k)}\odot (1-\lambda^{(k)}_t)+\lambda^{(k)}_t \tilde c_t,
\end{align}
where each $c_t^{(k)}$ is an exponentially moving average at time scale $\sim 1/\lambda^{(k)}_t$. Finally, the output will be
\begin{equation}\label{eq:c_EMA_n}
c_t= \sum_{k=1}^n w_k c^{(k)}_t,
\end{equation}

The simple $\alpha-$RNN \citep{alphaRNN}, which are simplified GRUs, share the same assumption of a single time scale per output dimension and thus can be generalized in the same way.
Let us show now how extend LSTMs with a forget gate \citep{gers2000learning}. Starting from their output $h_t$, one has 
\begin{align}
    h_t&=o_t \odot \sigma_h(c_t)\\
    c_t&=f_t \odot c_{t-1} +i_t \odot \tilde c_t,
\end{align}
where $o_t$, $i_t$, and $\tilde c_t$ are determined from the input $x_t$ and the previous output $h_{t-1}$  with learned weights and $\sigma_h$  is a nonlinear function. Writing  $f_t=1-\lambda_t$ makes it  obvious that the cell vector $c_t$ evolves in the same way as $y_t$ in Eq.\ \eqref{eq:GRU_EMA} if $i_t\simeq \lambda_t$.

Extending LSTMs to include $n$ time scales by cell state dimension is therefore straightforward: one needs to compute $n$ EMAs and their associated $\lambda$s as follows

\begin{align}
f^{(k)}_t&=\sigma(W^{(k)}_f x_t +U^{(k)}_f h_{t-1} + b_f^{(k)} )\\
i^{(k)}_t&=\sigma(W^{(k)}_i x_t +U^{(k)}_i h_{t-1} + b_i^{(k)} )\\
c^{(k)}_t&=f^{(k)}_t \odot c^{(k)}_{t-1} +i_t ^{(k)}\odot \tilde c_t,
\end{align}
where $\tilde c_t$ follows Eq.\ \eqref{eq:c_EMA_n}. Note that one could set $i^{(k)}_t = 1-f^{(k)}_t = \lambda^{(k)}_t$ and not learn the weights associated to $i_t$. Learning as well $i^{(k)}$ is equivalent to modulate the importance of the update, which is known as to as cognitive bias \citep{palminteri2017confirmation}: this is made clear by writing $i^{(k)}_t=v^{(k)}_t(1-f^{(k)}_t)=v^{(k)}_t\lambda_t^{(k)}$, where $v^{(k)}_t$ is the modulation of learning speed.

We will focus on the case $n=2$:
\eqref{eq:c_EMA_n} amounts to
\begin{equation}\label{eq:c_t_2exp}
    c_t= c^{(1)}_t \odot \alpha +(1-\alpha)\odot c^{(2)}_t,
\end{equation}
where the vector $\alpha$ is learnable  and its components are bounded to the $[0,1]$ interval. We call LSTMs with several time scales ($n>1$) per dimension VLSTMs, which stands for very long short term memory.

Note that LSTMs with a sufficiently large cell dimension $N_h$ can in principle learn to superpose timescales in the same way as Eq.\ \eqref{eq:c_EMA_n} by learning one time scale per dimension and using final dense layer to learn how to combine them. However, imposing constraints (or equivalently, injecting some known structure) is known to lead to faster learning and better results (e.g. Physics-guided deep learning, see \cite{thuerey2021pbdl} for a review). 

Naively, when learning to predict a process that is not too noisy, we expect the difference between VLSTM and LSTM to be the highest when $N_h=1$, i.e. precisely when LSTMs do not have the possibility to compute long-term averages and to decrease when $N_h$ increases.

\subsection{Volatility prediction}

\begin{figure}
    \centering
       \includegraphics[width=0.45\textwidth]{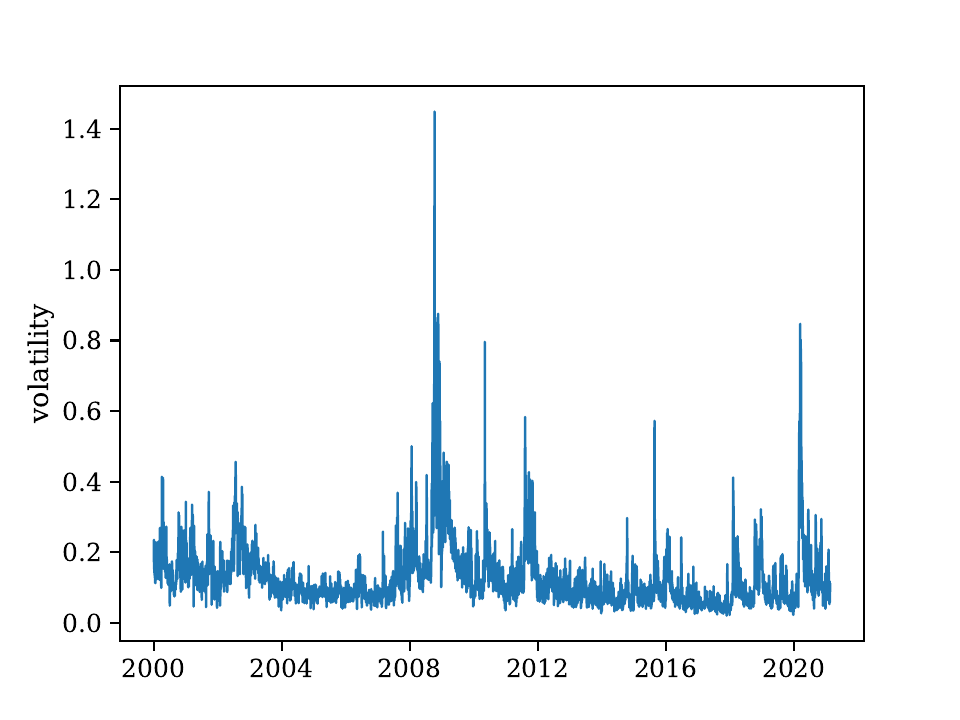}    
    \caption{Volatility of the SPX index (annualized) as a function of time, showing traces of long-term dependence. Data source: Oxford-Man Institute \citep{Heber:2009}.}
    \label{fig:log_vol_vs_t}
\end{figure}

Given an asset price $P_t$, its and its log return $r_t=\log P_t-\log P_{t-1}$, the asset price volatility $\sigma$ is defined as $\sigma^2=E(r^2)$. The dynamics of financial markets is ever-changing, which results in a temporal dependence of $\sigma$ with clear patterns of long-term dependence \citep{cont2001empirical} (see Fig.\ \ref{fig:log_vol_vs_t}).

Risk management, portfolio optimization. and option pricing benefit from the ability to predict the evolution of $\sigma_t$. Fortunately, $\sigma_t$ is relatively easy to predict, owing to its long memory \citep{cont2001empirical}: for example,  its auto-correlation decreases very slowly, presumably as a power-law over more than a year. Econometric models include GARCH, whose simplest version involves only one timescale, while many variations use several time scales \citep{zumbach2001heterogeneous,corsi2009simple,zumbach2015cross}. Rough volatility \citep{gatheral2018volatility}, on the other hand, considers $\log \sigma_t$ as fractional Brownian motion and thus includes all the time scales. As can be expected, rough volatility models outperform GARCH-based models for volatility prediction. Using LSTMs for volatility prediction is found e.g. in \cite{KIM201825,filipovic2021machine,rosenbaum2022universality} that use various types of predictors (including GARCH models) and architectures. Notably, \cite{rosenbaum2022universality} show that the average prediction of 10 stacked LSTMs with past volatility and price return as predictors match the performance of rough volatility.

\subsection{Architecture and hyperparameters}

Our first aim is to characterise the effects of multiple time scales per cell dimension. Therefore, we compare simple non-stacked LSTMs with or without the proposed modification. Stacked LSTMs can learn additional time scales at the cost of doubling the number of parameters, which we precisely wish to avoid here.
We pass the outputs $h_t$ of the LSTMs and VLSTMs through a dense layer of size $N_h$ with sigmoid activation functions, so as to combine the outputs in a non-trivial way, and a final dense layer with linear activation. Both final layers have a bias term, which allows the model to learn a baseline volatility level.

We report a systematic study of the relative performance of LSTMs vs VLSTMs. We vary the sequence length $T_\textrm{seq}$ from 10 to 100 by steps of 15, and the dimension of the hidden state $N_h\in\{1,\cdots,5\}$. Finally, we train models with and without biases (except for the final two dense layers which always have biases). There are thus 70 variations of hyperparameters per architecture choice.

For each hyperparameter and architecture couple, we train 20 networks which yields 2800 models altogether.  We use a standard 60/20/20 train/validation/test splits and apply early stopping criterion of the minimum validation loss over the 5 last epochs, with a maximum of 1000 epochs. Batch size is set to 128.   
We train the networks to predict $\log \sigma_{t+1}$ with an MSE loss function.

Data comes from volatility computed by Oxford-Man Institute from intraday data with the two-scale realized kernel estimation method \citep{barndorff2008designing}, which contain volatility time series for 31 indices and 2117 to 5385 data points per index \citep{Heber:2009}. Since the volatility individual time series start and end at heterogeneous dates, we used the dates to define the train/validation/test splits: the train set ranges from 2000-01-04 to 2012-09-06, validation set from 2012-09-07 to 2016-11-23 and test set from 2016-11-24 to 2021-02-17. This is necessary as the time series are cross-correlated, hence, splitting them according to their respective length would cause information leakage from the future and thus overfitting.

\section{Results}

\subsection{Average loss}
\begin{figure}
    \centering
    \includegraphics[width=0.45\textwidth]{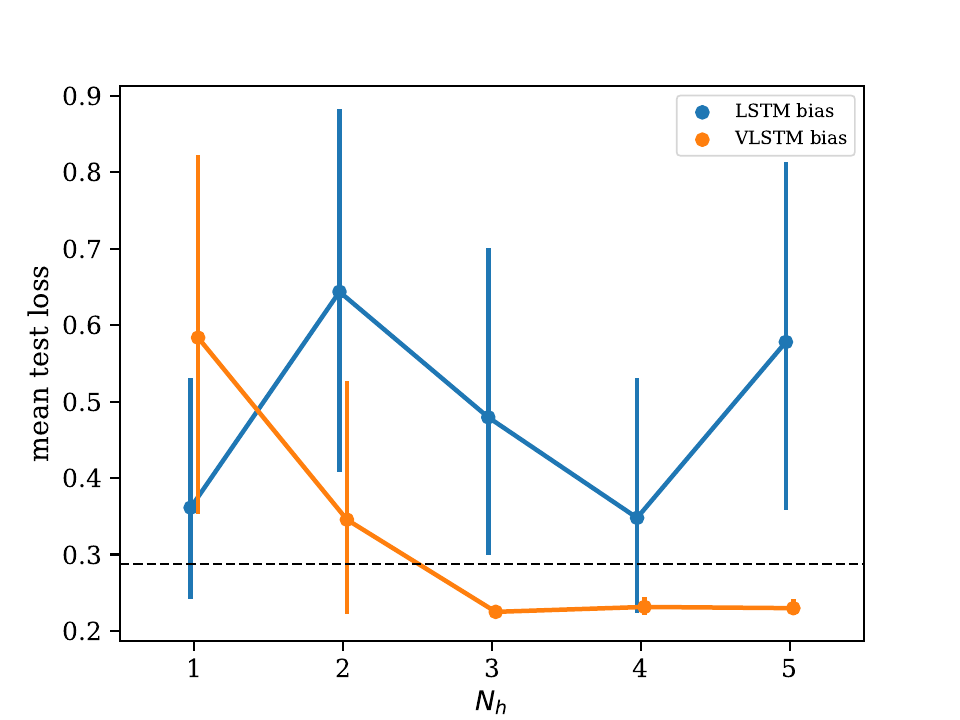}  
    \includegraphics[width=0.45\textwidth]{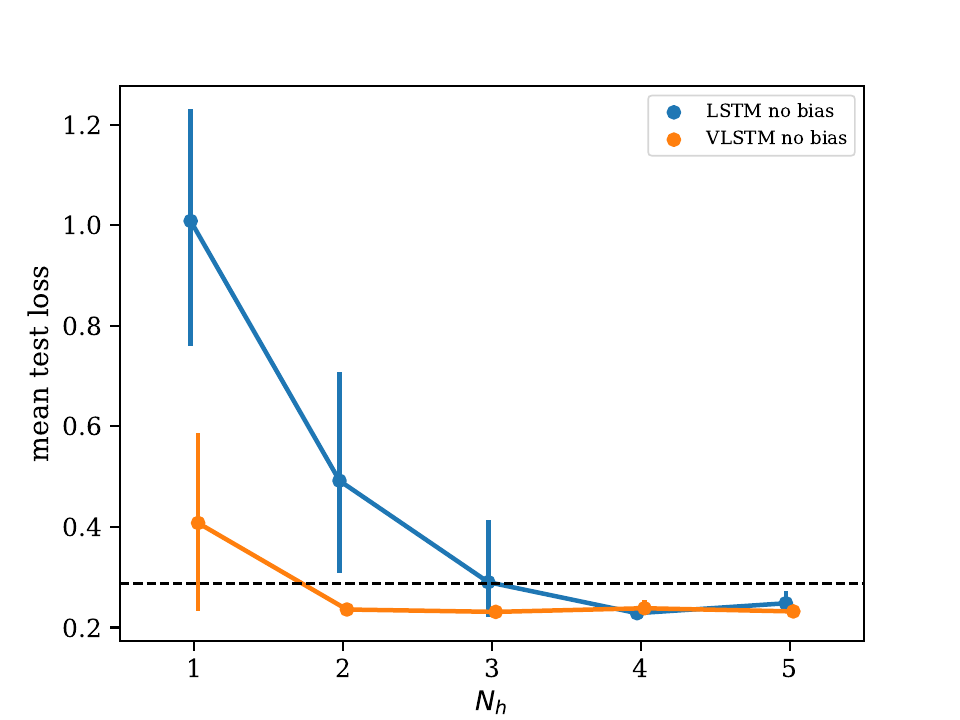}    
    \includegraphics[width=0.45\textwidth]{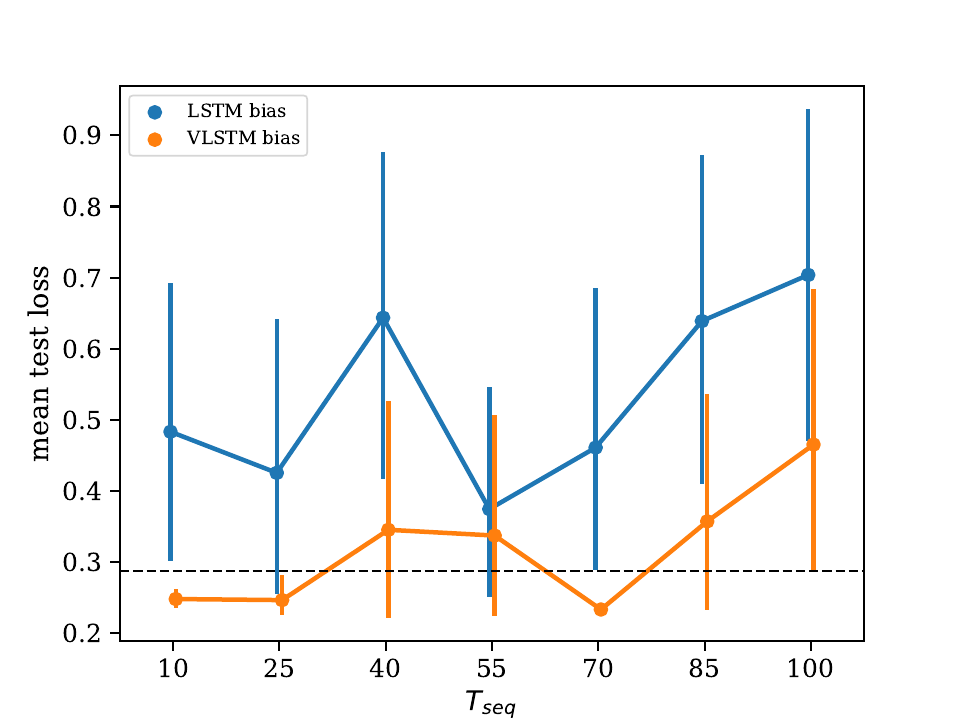}  
    \includegraphics[width=0.45\textwidth]{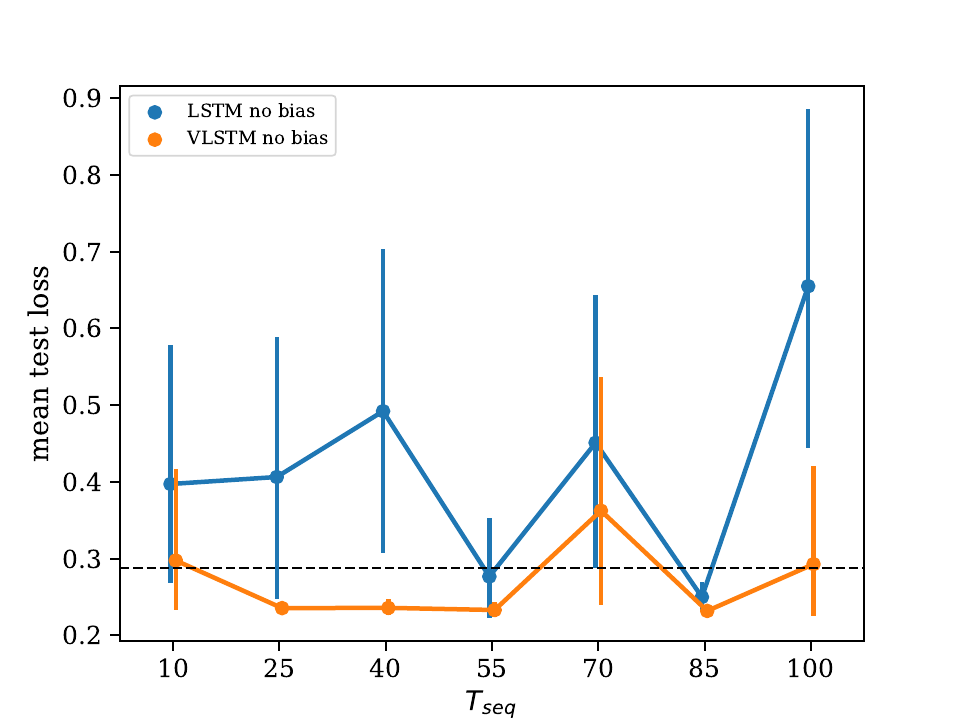}    
    \caption{Volatility prediction. Upper plots: mean test loss vs the memory cell dimension $N_h$ ($T_\textrm{seq}=40)$; lower plots: mean test loss vs the sequence length $T_\textrm{seq}$ ($N_h=2$). Left plots: (V)LSTMs with bias weights; right plots: (V)LSTMs with no bias weights.  The dashed line is the average MSE of predictions made with rough volatility models.}
    \label{fig:test_loss}
\end{figure}

Let us plot the average test loss of LSTMs and VLSTMs as a function of $N_h$ at fixed $T_\textrm{seq}$, the dimension of the memory cell, and of $T_{seq}$ at fixed $N_h$. This approach is taken by \cite{rosenbaum2022universality} who trained 10 LSTMs instead of 20 here. Figure\ \ref{fig:test_loss} shows that VLSTMs  enjoy a sizeable advantage on average. We note that when $N_h=1$, our initial intuition was correct: VLSTMs have a smaller average test loss for all variations of hyperparameters ($T_\textrm{seq}$ and bias)

Large loss fluctuations among models are associated with large average test losses for both VLSTMs and LSTMs; however test losses of VLSTMs are more likely to be small (and have accordingly small fluctuations). This is explained by a large difference in training convergence time, as shown below. We also note that, at least for volatility prediction, keeping bias terms in the computation of $i$, $f$, $\tilde{c}$, and $c$ (referred to as internal biases henceforth) is manifestly problematic; it turns out to be the default option both for PyTorch and Keras and is probably implicitly used in other papers. On the whole, we note that a simple average of the outputs of an ensemble of models leads to quite large fluctuations, hence that the question of the convergence of the models must be investigated and a way to select the good models would much improve the usefulness of LSTMs in that context.

\begin{figure}
    \centering
    \includegraphics[width=0.45\textwidth]{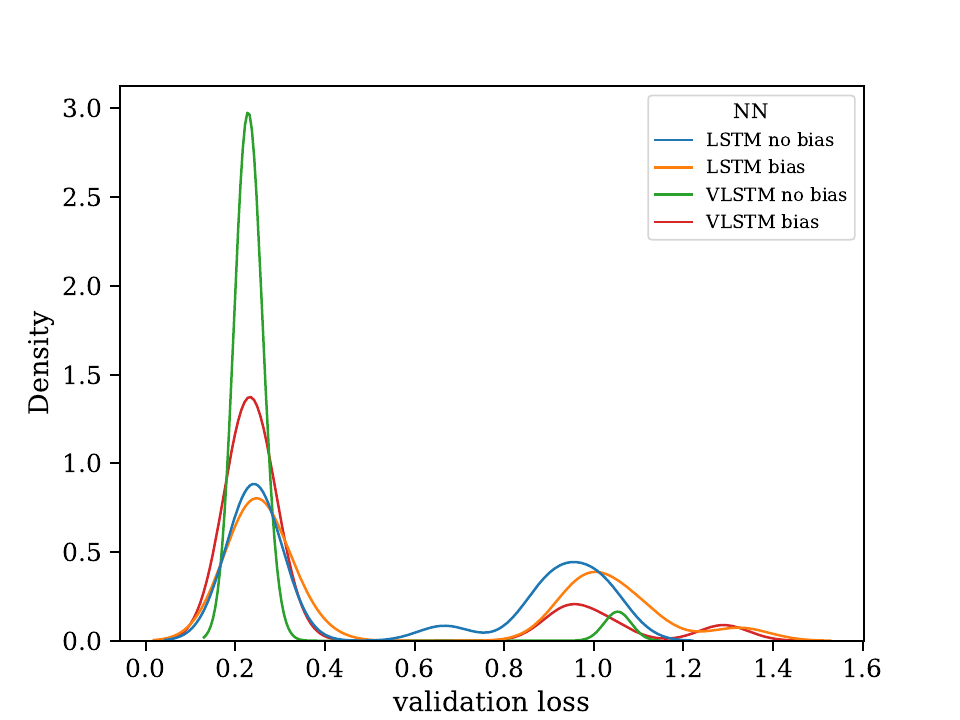} 
        \includegraphics[width=0.45\textwidth]{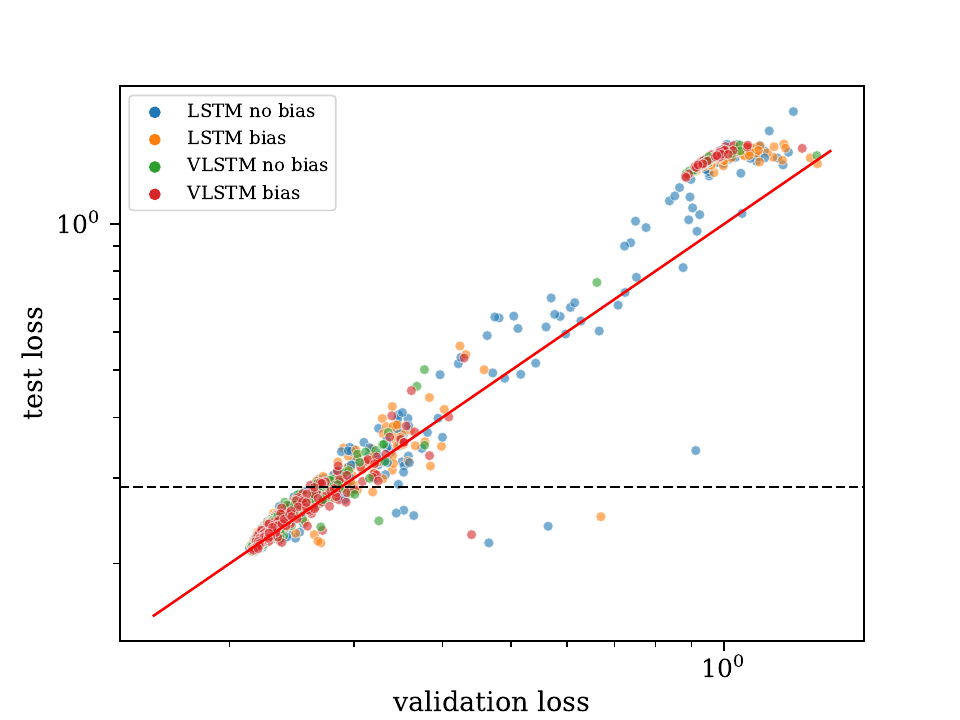} 
    \caption{Left plot: density of validation losses by architecture. Right plot: test loss vs validation loss. Multiple volatility time series prediction. The dashed line is the average MSE of predictions made with rough volatility models. }
    \label{fig:val_test_loss}

\end{figure}

Train convergence, it turns out, is a hit and miss process: some models are stuck in a high loss regime, while some models do learn a more realistic dynamical process and reach much lower losses. This yields a bi-modal density of losses (see Fig.\ \ref{fig:val_test_loss}). It is noteworthy that the fraction of VLSTMs that learn better is much larger. This is linked to the fact that VLSTMs learn much faster (see below) and that VLSTMs without internal biases are less likely to be stuck in a high loss regime.

Since the volatility process is well approximated e.g. by a rough volatility model \citep{gatheral2018volatility}, the test loss is commensurate with the validation loss as expected, itself commensurate with the train loss. We plot in Fig.\ \ref{fig:val_test_loss} the test loss versus the validation loss, which shows that test losses are accordingly also bimodal, with a majority of models not stuck in the high loss regime, some having a test loss smaller than rough volatility models.

\begin{figure}
    \centering
    \includegraphics[width=0.45\textwidth]{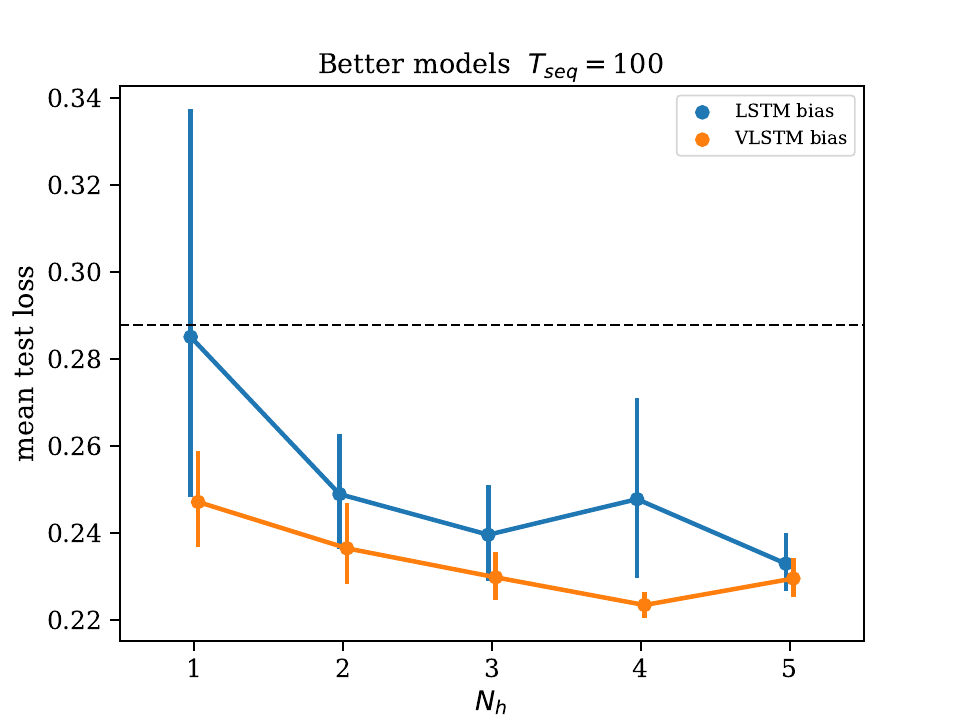}  
    \includegraphics[width=0.45\textwidth]{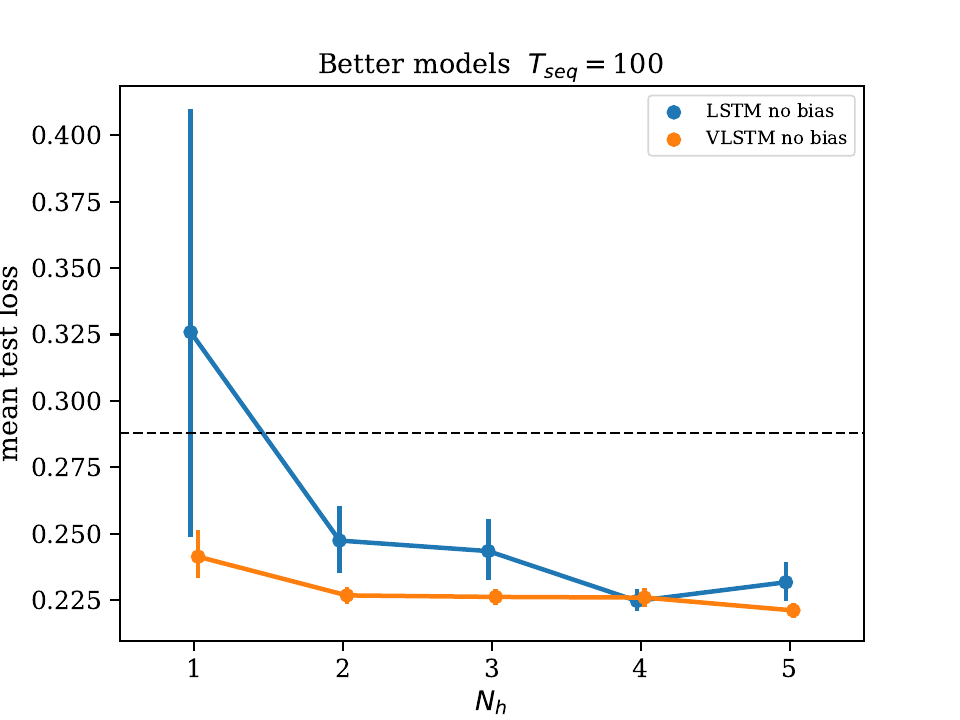} 
    \includegraphics[width=0.45\textwidth]{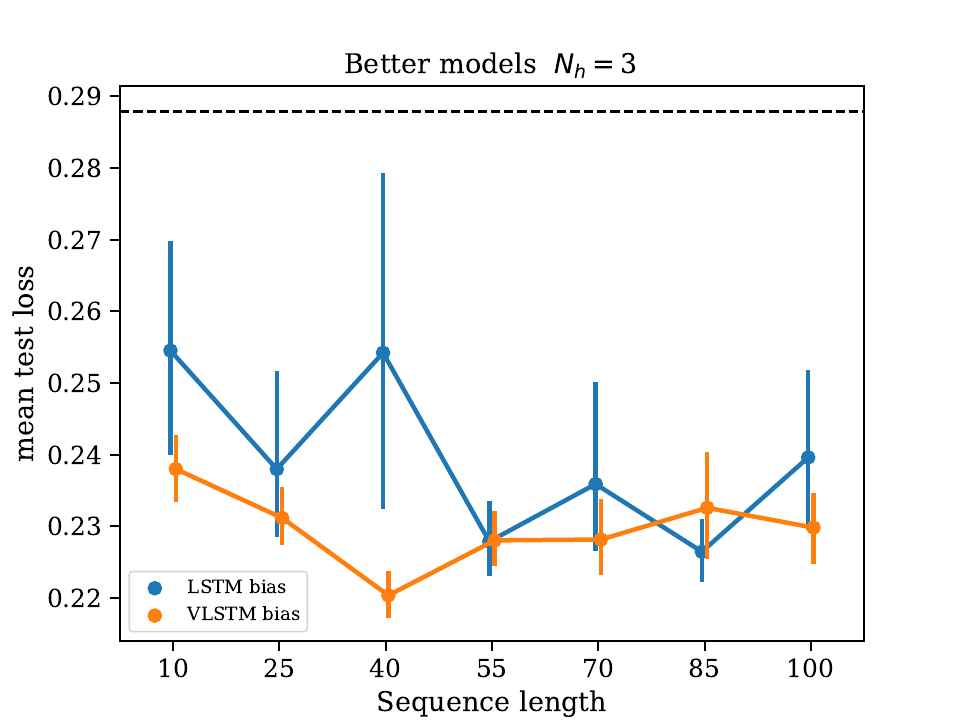}  
    \includegraphics[width=0.45\textwidth]{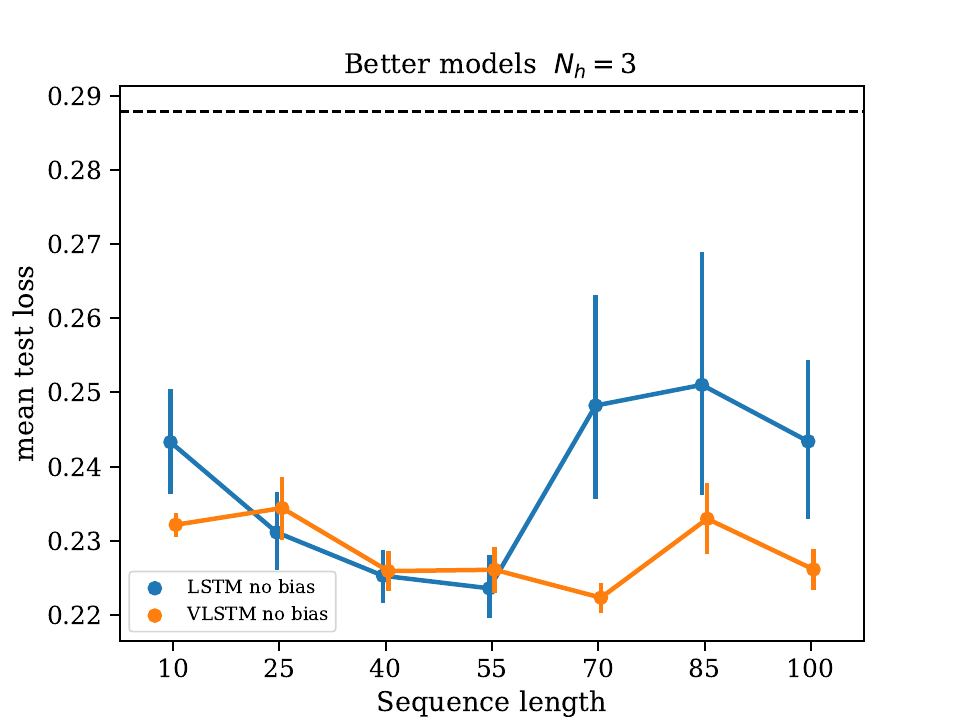}    
    \caption{Multiple volatility time series prediction test losses of the models with below-average validation losses. Upper plots: mean test loss vs the memory cell dimension $N_h$ ($T_\textrm{seq}=100)$; lower plots: mean test loss vs the sequence length $T_\textrm{seq}$ ($N_h=3$). Left plots: (V)LSTMs with bias weights; right plots: (V)LSTMs with no bias weights.  The dashed line is the average MSE of predictions made with rough volatility models. }
    \label{fig:cond_test_loss}
\end{figure}

\subsection{Keeping the better models}

This result suggests a way to select the good models, since the validation loss distribution is bimodal and since the test losses are roughly proportional to validation losses. To select models whose validation loss belongs in the lower peak, we compute 9 quantiles $q(p)$ with regular sequence of probabilities $p={0.1,\cdots,0.9}$, and keep the models whose validation loss is smaller than the quantile corresponding to the maximum change between quantiles, a simple yet effective way to find well separated peaks. We call these models the better ones in the following. This procedure allows a fairer comparison between LSTMs and VLSTMs. 

VLSTMs are still better than LSTMs, even for larger $N_h$. Figure\ \ref{fig:cond_test_loss} plots the average test loss of the models with below-average validation loss versus $N_h$ and the sequence length. The test losses are now much closer, but VLSTMs still retain a sizable advantage: their test losses are both lower on average and their fluctuations are much smaller.

\begin{table}[]
    \centering
\begin{tabular}{llrr}
 Architecture & bias  &test loss  &test loss \\
               & &average & std dev. \\ \hline
rough vol. & & 0.288 & \\
LSTM       & yes &0.241& 0.032 \\
LSTM    & no &0.245& 0.057 \\
VLSTM      & yes & 0.232& 0.017 \\
VLSTM   & no &0.230& 0.015\\
 & 
\end{tabular}
\caption{Standard deviation of the test losses of the better models computed over all the values of $N_h$ and $T_\textrm{seq}$.  Multiple time series volatility prediction}
    \label{tab:my_label}
\end{table}

\begin{figure}
    \centering
    \includegraphics[width=0.45\textwidth]{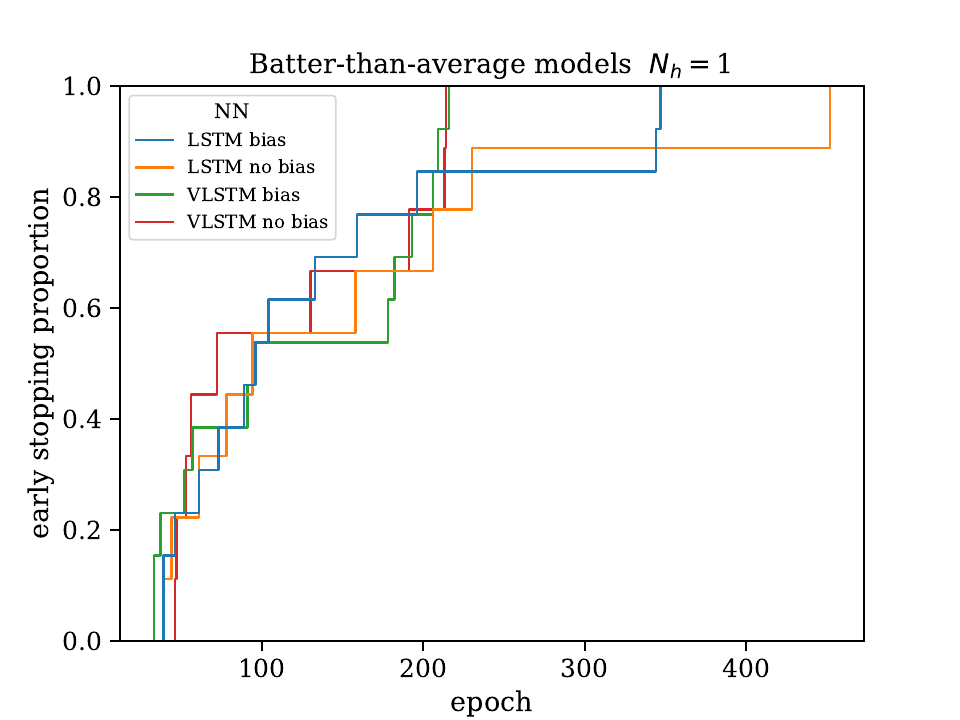}
    \includegraphics[width=0.45\textwidth]{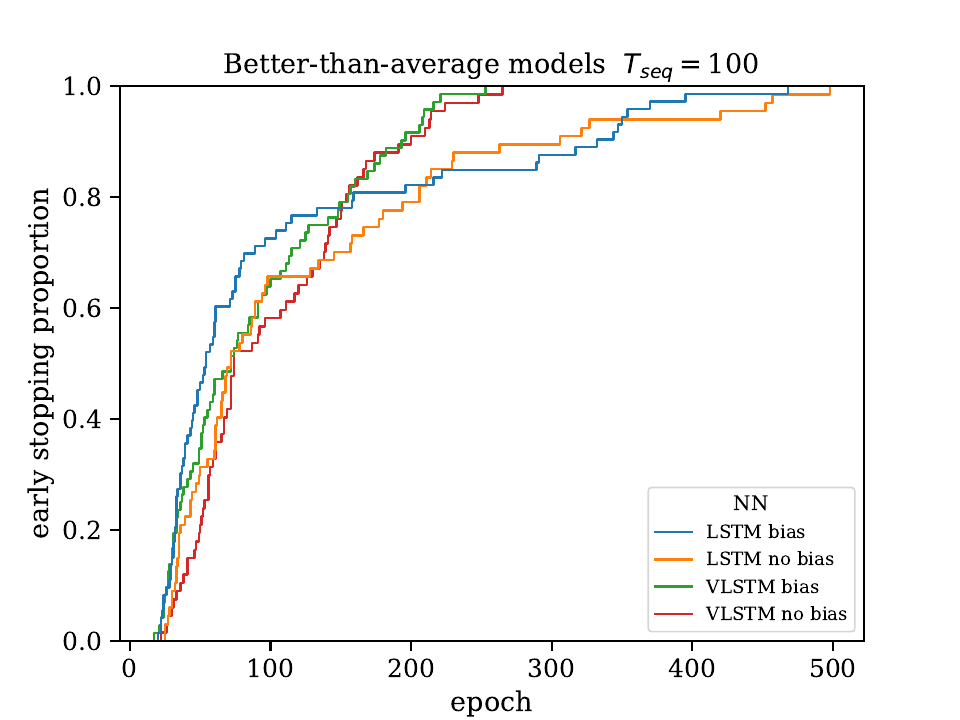}

    \caption{Fraction of models having converged before a given number of epochs. Left plot: $N_h=1$, right plot: all $N_h$;  multiple time series volatility prediction, $T_\textrm{seq}=100$.}
    \label{fig:time_to_learn}
\end{figure}

Both the variability of results and the strange results for $N_h=1$ when biases are allowed in the computation of the internal states of (V)LSTMs can be traced back to training convergence problems. A simple way to ascertain the main difference between VLSTMs and LSTMs is to measure the time it takes for their training to converge, i.e., to reach the early stopping criterion. Figure\ \ref{fig:time_to_learn} reports the fraction of models that have converged as a function of the number of epochs (limited to 1000). LSTMs need more epochs to be trained. We also found that the case $N_h=1$ and small $T_\textrm{seq}$ is hard to learn for this kind of architecture, the training of many models requiring more than 1000 epochs to reach the early stopping criterion.

\subsection{Best model}

\begin{figure}
    \centering
    \includegraphics[width=0.45\textwidth]    {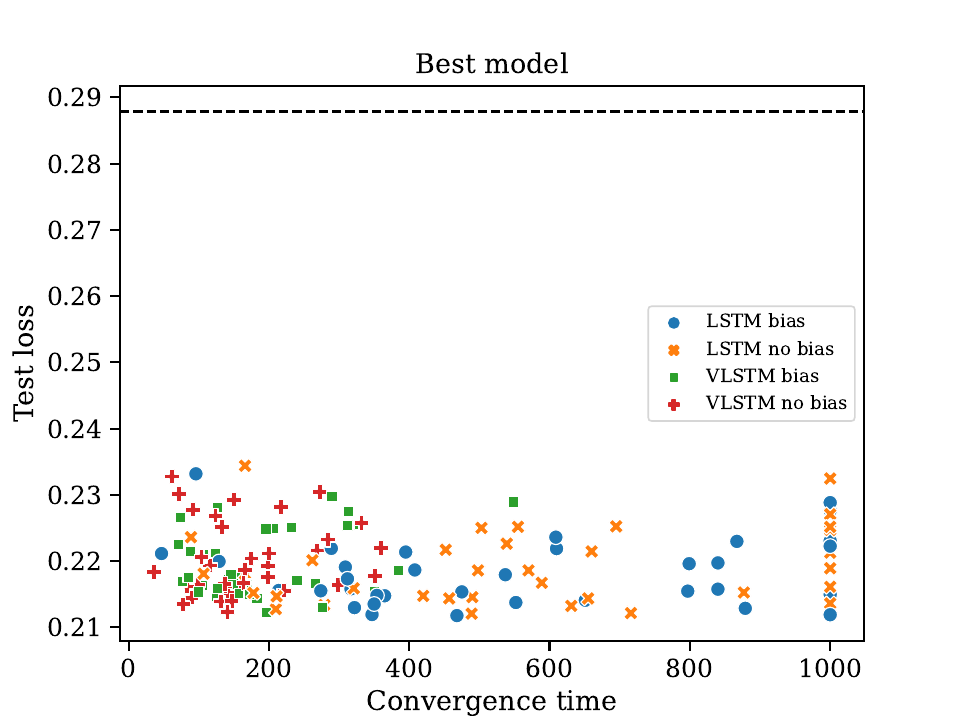}
\includegraphics[width=0.45\textwidth]{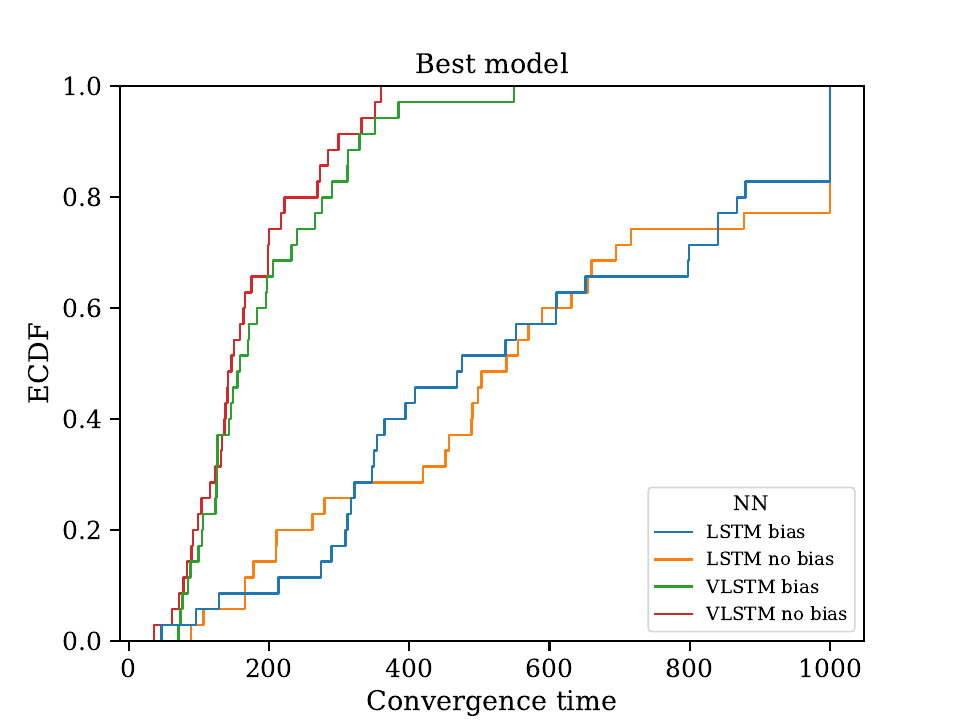}

    \caption{Left plot: test loss of the models with the best validation loss for all architecture and hyperparameter choices. Right plot: empirical cumulative distribution function of the convergence time for the four architecture choices. All values of $N_h$ and $T_\textrm{seq}$; multiple time series volatility prediction.  The dashed line is the average MSE of predictions made with rough volatility models.}
    \label{fig:testloss_convtime}
\end{figure}

Finally, let us investigate the test loss of the model with the best validation loss among the 20 models trained for each of the 140 hyperparameters/architecture choices. It turns out that under these conditions VLSTMs and LSTMs have essentially the same performance. What differentiate them however is the speed at which they learn. Let us plot the test loss versus the time of convergence for LSTMs and VLSTMs with and without biases (left plot of Fig.\ \ref{fig:testloss_convtime}): there is slight negative dependence between test loss and convergence times, the longer one learns, the better. Notably convergence times of LSTMs are spread all over the whole $[1,1000]$ interval, while VLSTMs converge before 400 epochs. The right plot of  Fig.\ \ref{fig:testloss_convtime} displays the ECDF of the convergence times, which shows a sizable difference between LSTMs and VLSTMs: whereas 20\% of LSTMs models do not manage to converge before 1000 epochs, all VLSTMs do before 400, except one, when biases are allowed.

Thus, training a given number of models is significantly shorter with VLSTMs because they do not need to learn how to approximate the kernel $K(x)$. One also sees that models with internal biases converge more slowly than those without them. We also wish to point out that because the fluctuation of validation losses among the trained models is much smaller for VLSTMs than for LSTMs, hence, that in practice, one needs to train fewer VLSTMs than LSTMs before finding a good one.  

\section{Conclusion}

Adding an explicit but flexible kernel structure to LSTMs brings significant improvements in every metric: number of epochs needed to reach convergence, overall prediction accuracy, and accuracy variation between models at fixed parameters. There is a cost as the number of trainable parameters of VLSTMs is larger than LSTMs are fixed hyperparameters, but doubling the number of time scales does not require to double the number of trainable parameters, thanks to the explicit kernel approximation structure. Although this paper focuses on LSTMs, the same idea can be applied to GRUs and $\alpha-$RNNs in a straightforward way. 

Our results mirror those of \cite{rosenbaum2022universality}: we also succeeded in training a single model at a time on many volatility time series of various  underlying asset types. This reflects the universality of volatility dynamics, a fact also hinted at by rough volatility and multi-scale GARCH-like models \citep{zumbach2015cross}. 

While it is hard to beat rough volatility for volatility prediction, we found that even simple LSTMs can beat it, provided that one trains several models and selects the best one according to its validation loss. Using LSTMs for that purpose requires to train more models over more epochs than using VLSTMs. Volatility prediction can be further improved by adding some more features, such as prior knowledge of predictable special events, and possibly by using more complex neural architectures.

\section{Code and data availability}

Full code, including the Keras VLSTM class, and data, are available at \url{https://github.com/damienchallet/VLSTM}.

\section*{Acknowledgments}
This work used HPC resources from the ``M\'esocentre'' computing center of CentraleSupélec and \'Ecole Normale Sup\'erieure Paris-Saclay supported by CNRS and R\'egion \^{I}le-de-France.

\bibliographystyle{plainnat}
\bibliography{references}

\end{document}